\documentclass[conference]{ieeeconf}

\IEEEoverridecommandlockouts
\usepackage{cite}
\usepackage{amsmath,amssymb,amsfonts}
\usepackage{graphicx}
\usepackage{textcomp}
\usepackage{xcolor}

\usepackage{amsthm}
\theoremstyle{definition}

\newtheorem{prob}{Problem}

\usepackage{siunitx}
\usepackage{lettrine}
\usepackage{epstopdf}
\usepackage{url}
\usepackage{epsfig}
\usepackage{graphicx}
\usepackage{algorithm}
\usepackage{algorithmic}
\usepackage{tabularray}
\usepackage[mathscr]{euscript}
\renewcommand{\baselinestretch}{0.91}
\renewcommand{\textfloatsep}{5mm} 

\newif\ifarxiv
\arxivtrue      

\begin{document}

\title{\bf Competitor-aware Race Management for Electric Endurance Racing}

\author{Wytze de Vries, Erik van den Eshof, Jorn van Kampen, Mauro Salazar
	\thanks{Wytze de Vries, Erik van den Eshof, Jorn van Kampen, and Mauro Salazar are with the Control Systems Technology Section, Department of Mechanical Engineering, Eindhoven University of Technology (TU/e), 5600 MB Eindhoven, The Netherlands (e-mail: w.a.b.d.vries@tue.nl; r.c.p.v.d.eshof@tue.nl; j.h.e.v.kampen@tue.nl; m.r.u.salazar@tue.nl).}}
\maketitle
\thispagestyle{plain}
\pagestyle{plain}
\begin{abstract}
Electric endurance racing is characterized by severe energy constraints and strong aerodynamic interactions. Determining race-winning policies therefore becomes a fundamentally multi-agent, game-theoretic problem. These policies must jointly govern low-level driver inputs as well as high-level strategic decisions, including energy management and charging. This paper proposes a bi-level framework for competitor-aware race management that combines game-theoretic optimal control with reinforcement learning. At the lower level, a multi-agent game-theoretic optimal control problem is solved to capture aerodynamic effects and asymmetric collision-avoidance constraints inspired by motorsport rules. Using this single-lap problem as the environment, reinforcement learning agents are trained to allocate battery energy and schedule pit stops over an entire race. The framework is demonstrated in a two-agent, 45-lap simulated race. The results show that effective exploitation of aerodynamic interactions is decisive for race outcome, with strategies that prioritize finishing position differing fundamentally from single-agent, minimum-time approaches. 
\end{abstract}

\section{Introduction}\label{sec:Intro}
\lettrine{T}{he} motorsports industry is rapidly electrifying. Hybrid powertrains are now standard in series such as Formula 1, while fully electric championships including Formula E and Extreme E compete at the highest level. A defining feature of electric racing is the severe energy limitation imposed by battery technology. As a result, performance depends critically on race management which includes the allocation of battery energy, powertrain operation, and charging decisions over the full race distance. Aerodynamic interactions between cars further complicate this task. Driving behind another car reduces drag and energy consumption on straights but compromises cornering performance due to reduced downforce. In energy-constrained races, strategic energy savings through slipstreaming can determine the final outcome~\cite{theRacePeloton}.

Unlike single-agent minimum-time racing, electric endurance racing requires optimal decisions under a fixed energy budget, with payoff defined solely by the finishing position relative to competitors. Because each competitor’s actions affect the feasible strategies of others, the problem is inherently game-theoretic. Success depends on balancing the energy saved while following against the cost of overtaking, ensuring optimal track position at the finish.

To address this challenge, we propose a bi-level optimization framework. The lower level determines the optimal in-lap driver inputs via a multi-agent optimal control problem that captures aerodynamic interactions. The upper level learns strategy engineer decisions made at the start of each lap, including energy allocation, pit-stop timing, and charging duration, to maximize finishing position.

\begin{figure}[t]
	\centering
\includegraphics[width=1\linewidth]{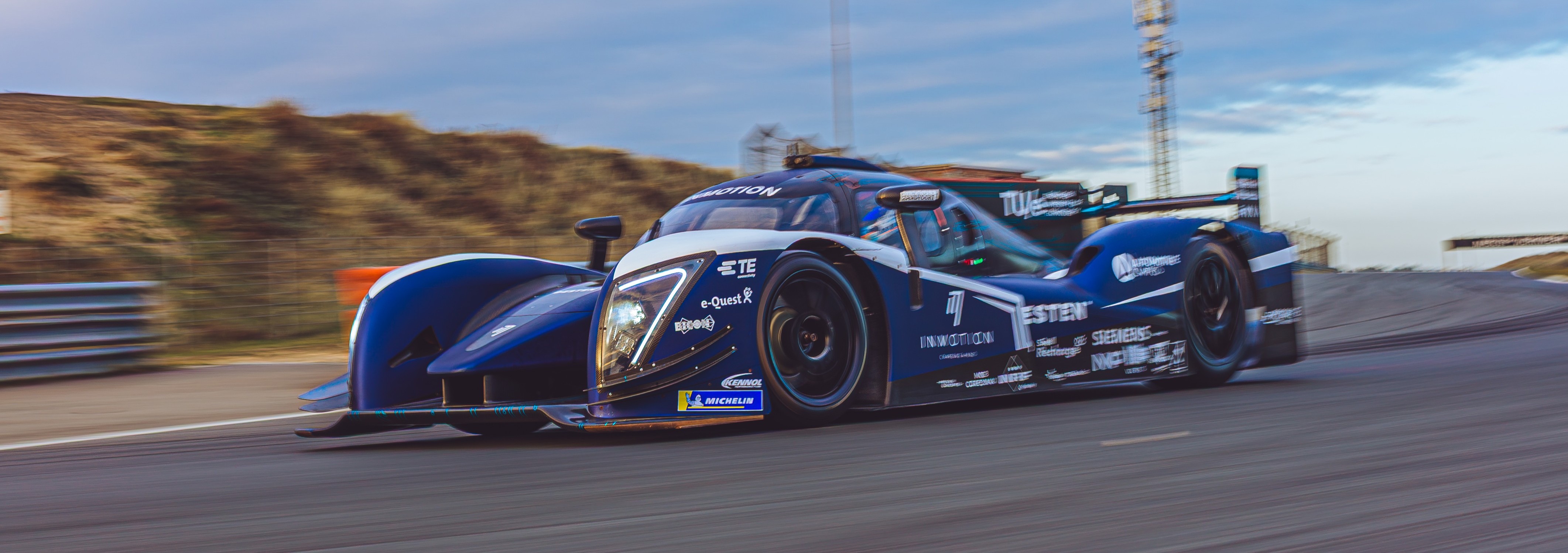}
	\caption{InMotion's fully electric endurance race car at the Zandvoort circuit.}
	\label{fig:revolution}
\end{figure}

\subsubsection*{Related Literature}
This work pertains to two research streams, lap time optimization and full-race strategy optimization.

Single-lap optimization determines the optimal racing line and powertrain operation to minimize lap time. Several approaches compute the racing line from geometry or measurements~\cite{Heilmeier_minCurveTraj, Salazar_OCS_Paper1}, followed by powertrain optimization along a fixed path~\cite{Salazar_OCP_Paper2, Borsboom_Convex_Framework_Electric_Race_Cars}. Joint optimization is typically addressed using nonlinear programming~\cite{Liu_FE_OptiLTC, limebeer2018dynamics}. While effective for single vehicles, these methods neglect realistic competitor interaction.

Game-theoretic extensions optimize multiple vehicles simultaneously~\cite{Burger_Interaction_Aware_Motion_Planning_BiLevelOpt, Dreves_Nash_Approach_OCP, Fieni_2D_CompetitorInteraction}, but generally assume cooperative behavior or selfish lap-time minimization. Furthermore, collision constraints are often imposed symmetrically on both agents \cite{Fieni_2D_CompetitorInteraction}, which can lead to unrealistic behavior in which the leading car yields its position to avoid a collision with an overtaking vehicle. In practice, however, the responsibility for avoiding contact typically lies with the overtaking car. Policy-based approaches, including iterative best response~\cite{Wang_Planning_Competetive_MultiVeh} and reinforcement learning~\cite{Wurman_GranTurismo}, capture competitive behavior but rely on short horizons, limiting their ability to enforce race-long energy constraints.

Full-race strategy optimizes decisions over multiple laps, including energy allocation, pit-stop timing, and charging duration. Traditionally, race simulations~\cite{Heilmeier_Race_Simulation, Bekker_F1Strat_Discrete_Simulation} heuristically compare strategies. More recent works formulates direct optimization problems for hybrid~\cite{Duhr_Minimum_Race_Time_Energy_Alloc_F1} and fully electric vehicles~\cite{vanKampen_MaximumDistanceRaceStrategies}, and apply machine learning methods such as Monte Carlo tree search and reinforcement learning~\cite{Liu_FE_Strategy_Reinforcement_Learning, FieniRL1}. However, these methods neglect competitor interactions.

In~\cite{vanKampen_MPC_CompetitorInteraction}, a MPC approach is presented, where competitors are treated as disturbances rather than strategic agents. Dynamic programming is used in~\cite{Aguad_PitStopF1_DP} to account for the active response of opponents but is limited to discrete actions.

Reinforcement learning approaches~\cite{Liu_FE_Strategy_multiCar, FieniRL2, FieniRL3} show potential for learning multi-agent race strategies. However, they typically rely on simplified models of inter-agent interaction, where the influence of other agents is represented mainly through the time gap. While the modeling of such interactions is the key distinction between single-agent and multi-agent formulations, reducing it to a time-gap-based lap time penalty overlooks important aspects. In practice, these interactions are more complex. For instance, the outcome and efficiency of an overtaking maneuver depend not only on the time gap, but also on factors such as energy deployment. A car that expends more energy can complete an overtake while incurring a smaller lap-time penalty, than one with a smaller energy advantage, a behavior that is not captured by these simplified models. Furthermore, these methods were applied to Formula E and Formula 1 races, which do not include charging decisions.

To the best of the authors knowledge, no existing framework simultaneously captures realistic multi-agent interactions and optimizes race-long energy and charging strategies in electric endurance racing.

\subsubsection*{Statement of Contributions}
In this work, we propose a bi-level, game-theoretic framework for competitor-aware race management in electric endurance racing. At the lower level, we formulate a game-theoretic multi-agent optimal control problem over a single lap. This problem optimizes the driver inputs, including powertrain operation and racing trajectories for minimizing relative position, while explicitly capturing aerodynamic interactions and asymmetric collision-avoidance constraints inspired by real-world motorsport. At the upper level, we address long-horizon race strategy by learning policies for energy allocation, pit-stop timing, and charging duration over the full race distance. These lap-wise decisions, made by strategy engineers at the start of each lap, are optimized using reinforcement learning.
The framework is demonstrated in a two-agent setting, showing that strategic exploitation of aerodynamic interactions is decisive for race outcome, and that competitor-aware race policies differ fundamentally from single-agent, minimum-time strategies.

\subsubsection*{Organization}
The remainder of this paper is organized as follows. We start by introducing the problem for a single agent and a single lap in Section~\ref{sec:LowLevel}, after which we extend the single-lap problem to multiple agents using game theory in Section~\ref{sec:GameTheory}. We then extend the problem to multiple laps in Section~\ref{sec:HighLevel}, where we introduce the reinforcement learning methods we used to train a policy to make high-level decisions about race strategies. In Section~\ref{sec:Results} we present the numerical results, and Section~\ref{sec:Conclusion} summarizes the conclusions and outlines directions for future research.

\section{Single-Lap Single-Agent}\label{sec:LowLevel}
This section formulates the single-lap optimal control problem for each agent to determine optimal driver inputs. The track and vehicle models build on established methods, while we introduce a new collision-avoidance formulation that enforces compliance with motorsport regulations. In addition, we propose a different objective function that promotes competitive behavior rather than purely selfish lap-time minimization.

\subsection{Vehicle Model}
\begin{figure}
	\vspace{2mm}
	\centering
	\includegraphics[width=1\linewidth]{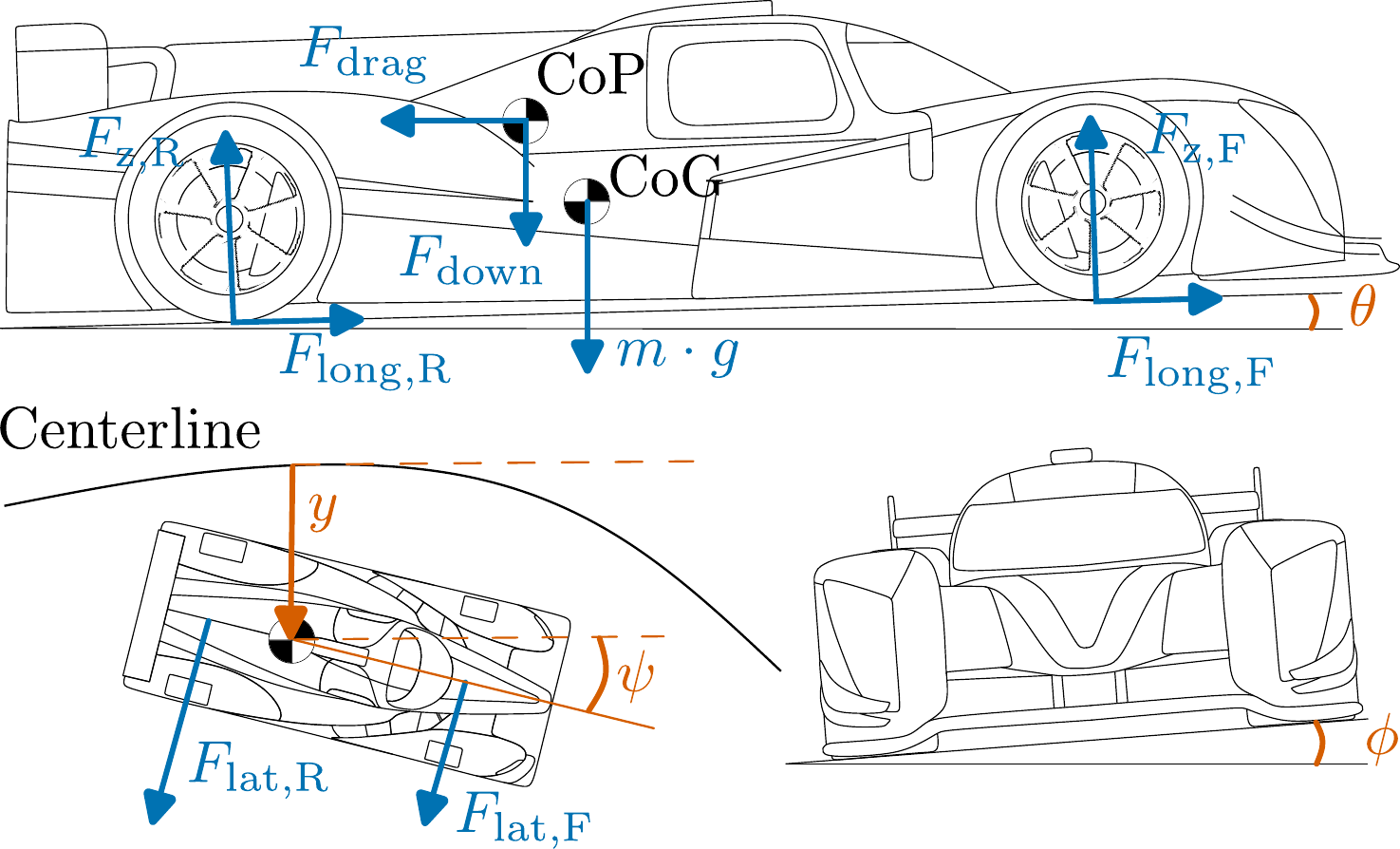}
	\caption{Free-body diagrams of the race car illustrating the longitudinal, lateral, and vertical force components shown in blue. Aerodynamic drag $F_{\mathrm{drag}}$ and downforce $F_{\mathrm{down}}$ act at the center of pressure (CoP), while gravitational force $m g$ acts at the center of gravity (CoG). Longitudinal and lateral tire forces at the front and rear axles are denoted by $F_{\mathrm{long},F}$, $F_{\mathrm{long},R}$, $F_{\mathrm{lat},F}$, and $F_{\mathrm{lat},R}$. Vehicle orientation (shown in orange) is described by the orientation $\psi$, and lateral displacement $y$ measured relative to the circuit center line. The slope and banking angles of the track are denoted by $\theta$ and $\phi$, respectively.}
	\label{fig:Frames}
\end{figure}

The problem is formulated in the spatial domain $s$, in which each agent controls a set of space-discretized inputs
$\mathbf{u}(s) = [F_\mathrm{em}(s),~F_\mathrm{brake,F}(s),~F_\mathrm{brake,R}(s),~F_\mathrm{lat}(s)]$.
Here, $F_\mathrm{em}$ denotes the electric motor output force, actuated by the driver via the accelerator pedal. The braking forces at the front and rear axles, $F_\mathrm{brake,F}$ and $F_\mathrm{brake,R}$ respectively, are applied through the brake pedal, while the lateral force $F_\mathrm{lat}$ is linked to the steering input. The vehicle dynamics are described by the state vector
$\mathbf{x}(s) = [v(s),~y(s),~\psi(s),~t(s),~\Delta p(s),~E_\mathrm{b}(s)]$,
where $v$ is the longitudinal velocity, $y$ the lateral displacement from the center line, $\psi$ the yaw angle deviation relative to the center line, $t$ the elapsed time, $\Delta p$ the position of the vehicle relative to its competitors, and $E_\mathrm{b}$ the remaining battery energy. An overview of the vehicle geometry and the forces acting on the car is shown in Fig.~\ref{fig:Frames}.  
The kinematic model is adapted from \cite{Fieni_2D_CompetitorInteraction}, with the inclusion of track slope and banking effects. The bicycle model is based on \cite{vanKampen_MaximumDistanceRaceStrategies} and extended to incorporate load-sensitive grip coefficients \cite{pacejka2012tire}. A quadratic loss model is employed for the powertrain, which is sufficient for the objectives of this study and the aerodynamic interactions were based on \cite{Fieni_2D_CompetitorInteraction}.

\subsection{Collision Avoidance Constraints}
This section introduces the collision avoidance constraints which depend on the relative positions of the vehicles. The longitudinal gap is expressed as a time-based spacing
\par\nobreak\vspace{-5pt}
\begingroup
\allowdisplaybreaks
\begin{small}
	\begin{equation}
			t_\mathrm{gap}(s) = t_i(s) - t_{-i}(s),
	\end{equation}
\end{small}%
\endgroup
\noindent where the subscript $i$ denotes the ego agent and $-i$ the opponent. The lateral gap is defined as the lateral distance between the two vehicles
\par\nobreak\vspace{-5pt}
\begingroup
\allowdisplaybreaks
\begin{small}
	\begin{equation}
		y_\mathrm{gap}(s) = y_i(s) - y_{-i}(s).
	\end{equation}
\end{small}%
\endgroup
\noindent To ensure safe interaction between agents, a minimum gap is enforced to prevent collisions. In most multi-agent formulations, this constraint is applied symmetrically to both agents at all times ~\cite{Fieni_2D_CompetitorInteraction, Wang_Planning_Competetive_MultiVeh}. In practice, such an approach does not accurately represent the rules of competitive wheel-to-wheel racing. In motorsport, the leading agent is generally free to select its racing line, while the following agent bears the responsibility of avoiding contact. If we apply the constraint to both agents, we could obtain solutions where the leading agent compromises their trajectory to avoid a collision, while in reality it would be the responsibility of the following agent.

To capture this asymmetry, the collision-avoidance constraint is activated or deactivated depending on the agents’ relative positions. The constraint is approximated using an elliptical shape defined as
\par\nobreak\vspace{-5pt}
\begingroup
\allowdisplaybreaks
\begin{small}
	\begin{equation}
	\label{eq:collision}
\left(\frac{t_\mathrm{gap}(s)}{t_\mathrm{gap,min}}\right)^2
+
\left(\frac{y_\mathrm{gap}(s)}{y_\mathrm{gap,min}}\right)^2
\geq 1 + \vartheta \cdot (\Delta p(s)-1),
	\end{equation}
\end{small}%
\endgroup
\noindent where $t_\mathrm{gap,min}$ is the minimum separation in the time domain, ${y_\mathrm{gap,min}}$ is the minimum lateral separation, and $\vartheta$ is a small positive constant. The right-hand side of~\eqref{eq:collision} effectively scales the size of the collision ellipse based on the relative position $\Delta p(s)$, ensuring that the following agent is subject to a larger safety envelope than the leading agent, while the small constant $\vartheta$ prevents the ellipses from collapsing at zero relative position when the agents are alongside each other.

Determining the relative position of the agents requires careful consideration, as it directly impacts the activation of the collision-avoidance constraint. Continuously evaluating the position based on the instantaneous time gap leads to numerical instabilities, since the responsibility for collision-avoidance could oscillate rapidly when the agents are in close proximity. To avoid this, the position is evaluated only at $J$ discrete points along the track and held constant between successive points. Specifically, the sampling locations are defined as
\par\nobreak\vspace{-5pt}
\begingroup
\allowdisplaybreaks
\begin{small}
	\begin{equation}
	0 = s_\mathrm{0} < s_\mathrm{1} < \dots < s_{J-1} < s_J = s_\mathrm{lap} ,
	\end{equation}
\end{small}%
\endgroup
\noindent and the relative position state is given by
\par
\begin{figure}
	\vspace{2mm}
	\centering
	\includegraphics[width=1\linewidth]{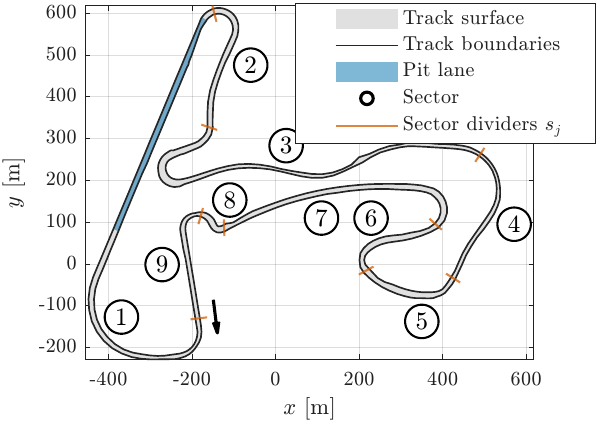}
	\caption{Map of the Zandvoort circuit showing the mini-sectors and the locations at which vehicle positions are evaluated for the upcoming mini-sectors. The lap starts and ends at the sector divider of the last corner before the pitlane, allowing us to model the pit dynamics at the start of the lap. The pit lane is indicated by the blue shaded area. Upon entering the pits, a speed limit of $60\,\mathrm{km\,h^{-1}}$ is enforced in this region.}
	\label{fig:trackmap}
\end{figure}

\nobreak\vspace{-5pt}
\begingroup
\allowdisplaybreaks
\begin{small}
	\begin{equation}
	\label{eq:pos}
\Delta p(s) = \tanh\big(k_\mathrm{pos} \cdot t_\mathrm{gap}(s_j)\big)~\text{for}~s \in [s_j, s_{j+1}),
	\end{equation}
\end{small}%
\endgroup
\noindent where $k_\mathrm{pos}$ is a tuning parameter, which results in the relative position being -1 when ahead, and 1 when behind. The measurement points are placed at the end of braking zones, as illustrated in Fig.~\ref{fig:trackmap}, reflecting common racing conventions where the car ahead at corner entry is entitled to select its preferred trajectory \cite{FIA_2025}.

\subsection{Single-lap, Single-agent Problem}

The agent’s objective is to minimize a weighted sum of its final relative position and lap time. The weighting parameter $w_\mathrm{pos}$ represents the lap-time-equivalent value assigned to a single position gain. Since the overarching objective in motorsport is to finish ahead of competing vehicles, regardless of absolute lap time, $w_\mathrm{pos}$ is chosen to be large such that relative position dominates the optimization. This reflects the fact that a slower lap is acceptable if it results in a net gain in relative track position at the end of the lap.

Nevertheless, lap time is retained in the objective. This ensures that the agent remains incentivized to drive efficiently when a position change is not achievable within the current lap. In such cases, minimizing lap time helps the agent stay close to competitors, thereby increasing the likelihood of creating or exploiting overtaking opportunities in subsequent laps. As a result, the objective balances the short-term competitive necessity of relative track position with the longer-term strategic value of maintaining race pace.

The resulting single-agent single-lap optimal control problem is therefore given by the following:
\begin{prob}[Single-agent single-lap problem]\label{prob:SingleLap1}
	The optimal competitive control strategy of each agent is the solution of
	\newlength{\mylength}
	\settowidth{\mylength}{\quad \text{Vehicle Dynamics:} \quad}
	\begin{equation*}
		\begin{aligned}
			\min_{\mathbf{x}(s),\mathbf{u}(s)} \quad & \Delta p(s_\mathrm{lap})+ \frac{1}{w_\mathrm{pos}} \cdot t(s_\mathrm{lap})  \\
			{\text{subject to:}}& \\
			&{\quad \text{Kinematics}} \\
			&{\quad \text{Bicycle model}}\\
			&{\quad \text{Powertrain model}} \\
			&{\quad \text{{Interactions}}} \\
			&{\quad \text{Boundaries}} \\
		\end{aligned}
	\end{equation*}
\end{prob}
\noindent with $s_\mathrm{lap}$ denoting the total distance of a lap.

\section{Single-Lap Multi-Agent}\label{sec:GameTheory}
This section introduces Nash and Stackelberg equilibrium concepts and discusses their interpretation in a racing context. We then describe our solution approach, reformulating the multi-agent problem via Karush-Kuhn-Tucker (KKT) conditions into a single NLP, and discuss how the single-lap solution is used as a state–transition map to extend the framework to multiple laps.

For clarity of notation, we denote the agents by $A$ and $B$, and when discussing a single agent, we use $i$ to denote the ego agent and $-i$ to denote the opposing agents.

\subsection{Nash and Stackelberg Equilibria}
A Nash equilibrium is defined as a set of actions in which no agent can improve its objective by unilaterally modifying its own action, while the actions of the other agents remain fixed. In our racing game, this means that no agent holds a strategical advantage over its opponents. \\
In contrast to Nash games, Stackelberg games introduce a hierarchical structure between the agents. One agent acts as the leader and commits to a strategy first, whereas the other agent, referred to as the follower, optimizes its response after observing the leader’s decision. The leader therefore anticipates the optimal reaction of the follower and incorporates this knowledge into its own optimization problem. Determining which agent assumes the role of leader or follower depends on the structure of the interaction. In sequential games, the leader is the agent that selects its action first, with the follower responding after observing this choice. However, in our racing game, agents act simultaneously, there is no inherent temporal ordering, as we optimize the whole lap simultaneously. In such cases, the leader can be interpreted as the agent that initiates a strategic action, such as initiating an overtaking maneuver, to which the opponent then reacts, for example with a blocking maneuver. 

We have selected Nash equilibria for our framework as it most accurately reflects our game where both agents are identical and have the same strategic options.

Solving the problem using optimization methods yields deterministic, open-loop trajectories for each agent. Rather than constructing stochastic policies, the framework solves for the decisions over the entire lap, resulting in locally optimal trajectories. The solution can therefore be interpreted as what the agents should have done if they had perfect hindsight.

In reality the problem is stochastic as agents do not know how their opponents will respond. However, the deterministic formulation provides a clear and reproducible description of locally optimal strategic behavior. It captures how the agents should allocate energy and select actions given perfect knowledge of the game, making it well suited for the extension to the multi-lap framework. 

We use the common method of using the KKT conditions to reformulate our problem into a single NLP. For further information about the mathematical formulation and implementation details we refer to \cite{Fieni_2D_CompetitorInteraction}.

\subsection{From Single-lap to Multi-lap} \label{subsec:single2multi}
In Section~\ref{sec:LowLevel} we obtained the single-lap, single-agent problem formulation, and in this section we introduced methods for computing multi-agent solutions. Using this single-lap multi-agent formulation, we extend this framework to multi-lap scenarios. 

We describe the lap-to-lap evolution using a state and an action: the state is the time gap $t_{\mathrm{gap}}(k)$ at the start of lap $k$, and the action is the per-lap battery allocation  $\Delta E_{\mathrm{b}}(k)$ that the strategy engineer decides at the start of each lap. These variables form the boundary conditions of Problem~\ref{prob:SingleLap1}. By solving the single-lap, multi-agent optimization problem for those boundary conditions, we obtain the state at the start of the next lap $t_{\mathrm{gap}}(k+1)$. In this work we limit ourselves to two agents, as the amount of possible combinations of boundary conditions grows exponentially with the number of agents.

Notably, the formulation is invariant to absolute time, as the single-lap dynamics depend only on the relative time gap between agents. This reflects the focus of this work on finishing position rather than absolute race time, and allows the multi-lap evolution to be fully described by the time gap alone.
Because of this invariance, the output of the single-lap, multi-agent problem yields a compact state transition of the form
\par\nobreak\vspace{-5pt}
\begingroup
\allowdisplaybreaks
\begin{small}
\begin{equation}\label{eq:tgap_general}
	t_{\mathrm{gap}}(k+1)
	\;=\;
	\rho\big(\Delta E_{\mathrm{b,A}}(k),\, \Delta E_{\mathrm{b,B}}(k),\, t_{\mathrm{gap}}(k)\big),
\end{equation}
\end{small}%
\endgroup
\noindent where $\rho$ denotes the mapping produced by solving the single-lap, multi-agent optimization problem. Note that in the remainder of this paper we define the time gap from the perspective of
\par\nobreak\vspace{-5pt}
\begingroup
\allowdisplaybreaks
\begin{small}
\begin{equation}
	t_{\mathrm{gap}}(k) = t_\mathrm{B}(k) - t_\mathrm{A}(k).
\end{equation}
\end{small}%
\endgroup

\section{Multi-Lap Multi-Agent}\label{sec:HighLevel}
In this section, we present the real-world multi-lap game we aim to solve. We then describe how this game and its agents are modeled, followed by an explanation of how we train a reinforcement-learning policy to maximize the competitiveness of the agents. 

In our game, two teams consisting of a driver and a strategy engineer compete to win a race consisting of a fixed number of laps $N_\mathrm{laps}$. At the start of each lap, each strategy engineer chooses how much battery energy to deploy for that lap, whether they enter the pits at the start of the lap to recharge the battery, and if they enter the pits how much battery energy they want to charge. Given these decisions, the single-lap multi-agent optimization problem determines how the time gap between the drivers evolves over the course of the lap. For each team, hereafter referred to as an agent, the objective is to finish the race ahead of its opponent. In this work, tire wear and tire selection are neglected. This assumption is based on the fact that tires can be changed during charging stops, and that the lifespan of endurance racing tires typically exceeds the distance a car can cover between charges.

The KKT-reformulation method that we apply to the single-lap problem is not able to solve this multi-lap problem as it contains, due to charging stops, non-smooth dynamics and binary variables. In fact, most multi-agent optimization methods \cite{IBR, OGDA, McMahan_DO, Adam_DO} struggle with the nonlinear, non-smooth and zero-sum nature of the problem considered here. A further challenge arises from the strategic nature of the interaction itself. In competitive multi-agent settings, deterministic strategies can lead to cyclic behavior analogous to the rock–paper–scissors game: for any deterministic strategy adopted by one agent, there exists a counter-strategy that results in a win for the opponent. As a result, purely deterministic policies fail to converge. To avoid this issue and to capture mixed-strategy equilibria, agents must instead adopt stochastic policies. Reinforcement learning \cite{MARLbook} has shown the most promise for computing approximate solutions in such settings. 

We therefore adopt a reinforcement learning framework and in the following section we will describe the models used to represent the environment and the agents. The reinforcement learning framework refers to the dynamics of our \emph{environment}, the \emph{actions} available to the agent, what the agents can \emph{observe} from the environment and how the agents are \emph{rewarded} for their actions. 

\subsubsection{Actions}
The first action is how much battery energy to expend during the upcoming lap $\Delta E_\mathrm{b}$. Allowing the agents to choose a raw energy value simultaneously would lead to unrealistic behavior as the trailing agent could inadvertently overtake the leader simply because the leader selected a lower energy expenditure than anticipated. In practice, the follower continuously observes the leader and adapts its inputs accordingly, making such accidental overtakes unrealistic. Modeling the interaction as a fully sequential game is also inappropriate, as it grants the follower complete knowledge of the leader’s planned action before making its own decision, which is also not representative of the real world.
To avoid this unrealistic behavior, we instead let the following agent select a change in time gap $\Delta t_{\mathrm{gap,target}}$ it intends to achieve at the end of the lap. This reflects whether the agent aims to remain behind or attempt an overtake. This structure preserves realistic observability and intent modeling while preventing unrealistic overtaking behavior.
We constrain the battery energy allocation per lap of the leader to
\par\nobreak\vspace{-5pt}
\begingroup
\allowdisplaybreaks
\begin{small}
\begin{equation}
	\Delta E_\mathrm{b,min} \leq \Delta E_\mathrm{b} \leq \Delta E_\mathrm{b,max},
\end{equation}
\end{small}%
\endgroup
\noindent where $\Delta E_\mathrm{b,min}$ and $\Delta E_\mathrm{b,max}$ are the lower and higher bound respectively. This is to prevent the energy allocation of the leader collapsing to zero when the agent does not want to lead the race. We do not constrain the energy allocation or time gap target of the following agent. 

The remaining actions concern the recharging strategy, which we define in terms of two components: the energy to be recharged during a pit stop, $E_\mathrm{b,charge}$, and the battery energy threshold for pit entry, $E_\mathrm{b,pit}$. The agent will enter the pit to recharge the battery if the battery state of charge is below the pit entry threshold $E_\mathrm{b,pit}$ at the start of the lap. Directly learning these absolute actions is a highly complex task. To simplify the problem, we leverage the single-agent minimum-time solution $\zeta$ from \cite{vanKampen_MaximumDistanceRaceStrategies}:
\par\nobreak\vspace{-5pt}
\begingroup
\allowdisplaybreaks
\begin{small}
\begin{equation}\label{eq:single-agent-sol}
	\begin{bmatrix}
		\hat{E}_\mathrm{b,charge}(k) \\
		\hat{E}_\mathrm{b,pit}(k) \\
		\hat{t}_\mathrm{rem}(k)
	\end{bmatrix} = \zeta(E_\mathrm{b}(k), N_\mathrm{laps}-k),
\end{equation}
\end{small}%
\endgroup
\noindent which returns the optimal values of $E_\mathrm{b,charge}$, $E_\mathrm{b,pit}$ and the estimated time it takes to complete the remaining laps, given the current battery energy $E_\mathrm{b}$ and the remaining number of laps $N_\mathrm{laps}-k$. Note that this solution does not account for the presence of other agents. The agent’s action then specifies deviations from this reference strategy, $\Delta E_\mathrm{b,charge}$ and $\Delta E_\mathrm{b,pit}$, such that the executed actions are
\par\nobreak\vspace{-5pt}
\begingroup
\allowdisplaybreaks
\begin{small}
\begin{equation}
	E_\mathrm{b,charge}(k) =  \hat{E}_\mathrm{b,charge}(k) + \Delta E_\mathrm{b,charge}(k),
\end{equation}
\end{small}%
\endgroup
\par\nobreak\vspace{-5pt}
\begingroup
\allowdisplaybreaks
\begin{small}
\begin{equation}
	E_\mathrm{b,pit}(k) =  \hat{E}_\mathrm{b,pit}(k) + \Delta E_\mathrm{b,pit}(k).
\end{equation}
\end{small}%
\endgroup
\noindent This formulation allows the agent to focus on making corrections to an existing strategy in order to account for the opponent, rather than learning the strategy from scratch. A similar approach is leveraged in \cite{FieniRL2} for Formula 1 racing.

\subsubsection{Observations}
The observations available to the agents are $\mathscr{O}(k) = [E_\mathrm{b}(k), \Delta p(k), t_\mathrm{gap}(k), \hat{E}_\mathrm{b,pit}(k),\hat{E}_\mathrm{b,charge}(k), N_\mathrm{laps}-k, \Delta n_\mathrm{pits}(k),  \hat{E}_\mathrm{b,opp}(k)]$, where $N_\mathrm{laps}-k$ is the number of laps remaining, $\Delta n_\mathrm{pits}$ is the difference in number of pit stops taken compared to the opponent, $,\hat{E}_\mathrm{b,charge}$ is an estimate of the remaining battery energy of the opponent. Since perfect information about an opponent’s battery state is unrealistic and not publicly available, the agent is instead provided with an estimate of the opponent’s remaining energy. In practice, this reflects the fact that agents can infer energy usage from observable behavior, such as the opponent’s pace on track and the duration of their pit stops, allowing them to deduce whether the opponent has been pushing aggressively or conserving energy.


\subsubsection{Environment}
The environment is mainly defined through the state-transition model which was introduced in Section \ref{subsec:single2multi}. To accelerate training, we replace this optimization with a neural-network-based surrogate model. The pit-stop influence on the time gap is modeled by accounting for both the charging duration required to recharge the energy and the time loss incurred while traversing the pit lane at reduced speed.

\subsubsection{Reward Design}
The sole objective for our agents is to win the race by finishing ahead of their opponent. Therefore we define the \emph{sparse reward} which is only applied at the end of the race as
\par\nobreak\vspace{-5pt}
\begingroup
\allowdisplaybreaks
\begin{small}
\begin{equation}
	r_\mathrm{sparse}(k+1) = w_\mathrm{win} \cdot \textnormal{sign} (t_\mathrm{gap}(k+1)) + w_\mathrm{gap} \cdot t_\mathrm{gap}(k+1),
\end{equation}
\end{small}%
\endgroup
\noindent where $w_\mathrm{win}$ and $w_\mathrm{gap}$ are weighting factors. We retain the magnitude of the time gap as it is a strong indicator for the agent whether it is moving in the right direction. Providing a reward only at the end of the race makes learning computationally difficult due to the temporal credit assignment problem, as it is unclear which actions during the race contributed to the final outcome. To improve training stability, we therefore provide a \emph{dense reward} throughout the race.
The dense reward is based on a prediction of the time gap at the finish. Using the free-stream solution introduced in \eqref{eq:single-agent-sol}, we reuse the remaining-time estimate $\hat{t}_\mathrm{rem}$ to predict the expected time gap at the end of the race as
\par\nobreak\vspace{-5pt}
\begingroup
\allowdisplaybreaks
\begin{small}
\begin{equation}
	\hat{t}_{\mathrm{gap,finish}}(k+1)
	= t_{\mathrm{gap}}(k+1)
	+ \hat{t}_{\mathrm{rem,B}}(k+1)
	- \hat{t}_{\mathrm{rem,A}}(k+1).
\end{equation}
\end{small}%
\endgroup
\noindent Directly applying a dense reward based on this prediction would change the objective and bias the learned policy away from the true objective of winning the race. In particular, the agent learns to optimize intermediate predictions rather than the final race outcome. Therefore, we incorporate it using \emph{potential-based reward shaping} (PBRS)\cite{PBRS} instead. PBRS includes such intermediate information while preserving the original objective. By defining a scalar potential function of the current state, PBRS guarantees that the optimal policy of the original problem remains unchanged. This allows the agent to focus learning on the sparse terminal reward that determines the race outcome, while still providing dense feedback that accelerates training.
We define the potential function as
\par\nobreak\vspace{-5pt}
\begingroup
\allowdisplaybreaks
\begin{small}
\begin{equation}
	\Phi(k+1) = w_\mathrm{pred} \cdot \varsigma\!\left(\hat{t}_{\mathrm{gap,finish}}(k+1)\right),
\end{equation}
\end{small}%
\endgroup
\noindent where $w_\mathrm{pred}$ is a weighting factor, and $\varsigma(\cdot)$ is a leaky saturation function,
\par\nobreak\vspace{-5pt}
\begingroup
\allowdisplaybreaks
\begin{small}
\begin{equation}
	\varsigma(x) = \textnormal{sign}(x) \cdot \,\log\!\left(1 + |x|\right),
\end{equation}
\end{small}%
\endgroup
\noindent which maintains sensitivity to changes in the predicted time gap while preventing excessively large shaping rewards.
The dense shaping reward is then given by
\par\nobreak\vspace{-5pt}
\begingroup
\allowdisplaybreaks
\begin{small}
\begin{equation}
	r_\mathrm{dense}(k+1)
	= 
	\left(
	\gamma\,\Phi(k+1) - \Phi(k)
	\right),
\end{equation}
\end{small}%
\endgroup
\noindent where $\gamma$ is the \emph{discount factor}. This formulation rewards actions that improve the predicted finish time gap relative to the opponent, while accounting for the discounted future value of such improvements. The final reward is defined as
\par\nobreak\vspace{-5pt}
\begingroup
\allowdisplaybreaks
\begin{small}
\begin{equation}
	r(k+1) =
	\begin{cases}
		r_\mathrm{dense}(k+1), & \text{if } k < N_\mathrm{laps}, \\[6pt]
		r_\mathrm{sparse}(k+1), & \text{if } k = N_\mathrm{laps}.
	\end{cases}
\end{equation}
\end{small}%
\endgroup

\subsubsection{Reinforcement Learning Process}
Proximal Policy Optimization (PPO)~\cite{PPO} is adopted due to its proven performance and robustness. A central challenge in multi-agent reinforcement learning is the non-stationarity of the environment. When two agents are trained directly against each other, or when a single agent is trained via self-play, it will likely fail to converge because the opponent’s policy is continuously changing. Training against a pool of fixed opponents is shown to significantly improve stability in competitive settings~\cite{OPENAI_policy_pool}, and therefore we use a similar approach. During training, the learning agent competes against a pool of opponent policies. At regular intervals, training is paused and the current policy is evaluated against all opponents in the pool. If the policy is able to consistently beat every opponent, it is added to the pool. Opponents are selected during training with a probability proportional to their win rate against the learning agent, such that stronger opponents are encountered more frequently while weaker opponents are still sampled occasionally. This prevents overfitting to only the stronger opponents. Training continues until the agent is no longer able to beat all opponents in the pool. The agent contains a common base with separate heads for each action. Furthermore, we employ the hyper-parameter tuning method introduced in~\cite{PolicyCollapse} to stabilize the training process. \\\\

\section{Results}\label{sec:Results}
This section presents numerical results for both the single-lap multi-agent and the multi-lap multi-agent problem. We base our use case on the rear-wheel-driven electric endurance race car of InMotion, shown in Fig.~\ref{fig:revolution}, driving at the Zandvoort circuit. We first show the results for the single-lap multi-agent problem, highlighting an overtaking maneuver to showcase the proposed collision-avoidance constraints. Animations of the results are available at \cite{YoutubeVideo}. We then continue with a discussion on the obtained multi-lap multi-agent policy, trained to execute a 45-lap race.

The single-lap problem was parsed with CasADi \cite{casadi} and solved using IPOPT \cite{ipopt}. Each agent was warm-started with their single agent solution of Problem~\ref{prob:SingleLap1}. The reinforcement learning framework was implemented using the Reinforcement Learning Toolbox from MATLAB \cite{RLToolbox}. The training procedure described in Section~\ref{sec:HighLevel} was continued until the learned policy could no longer consistently outperform the pool of opponent policies. At that point, the final policy was selected and evaluated in a full race simulation. The resulting policy is stochastic, with each action represented by a probability distribution characterized by a mean and standard deviation. In this section, actions are chosen deterministically by selecting the most likely value of each distribution.
\begin{figure}[t]
	\vspace{2mm}
	\centering
	\includegraphics[width=1\linewidth]{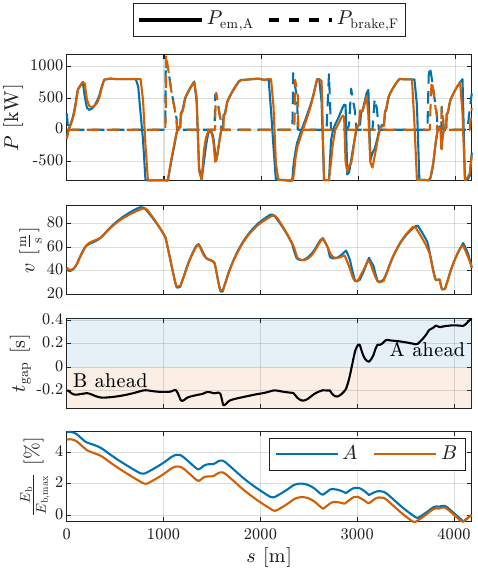}
	\caption{Electric motor power, front braking power, velocity, time gap, and battery energy usage for a Nash equilibrium on the Zandvoort circuit. The initial time gap is $0.2\,\mathrm{s}$ in favor of agent~B. The maximum battery energy usage per lap is $4.8\%$ and $5.3\%$ for agents~A and~B, respectively. Owing to drag reduction, agent~A achieves higher speeds and lower energy consumption in the first part of the lap. This energy advantage is subsequently exploited in the sequence of Turns~9 and~10, where agent~A overtakes its opponent.}
	\label{fig:singlelap}
\end{figure}

\subsection{Single-Lap Multi-Agent}
In this section, we present a single-lap multi-agent solution in which the initial time gap between the agents is set to 0.2s. The leading agent~B is assigned a battery energy allocation of 4.8\%, while the following agent~A is allocated 5.3\%. The resulting motor power, braking power, velocity profiles, battery energy trajectories, and inter-agent time gap are shown in Fig.~\ref{fig:singlelap}. Owing to the reduced aerodynamic drag in the slipstream of the leading vehicle, agent A achieves higher top speeds on the straights. This results in increased energy recovery during braking phases, further amplifying its energy advantage over agent B. The accumulated energy advantage is subsequently exploited to execute an overtake in the sequence of Turns~9 and~10.

\begin{figure*}[t]
	\centering
	\includegraphics[width=1\linewidth]{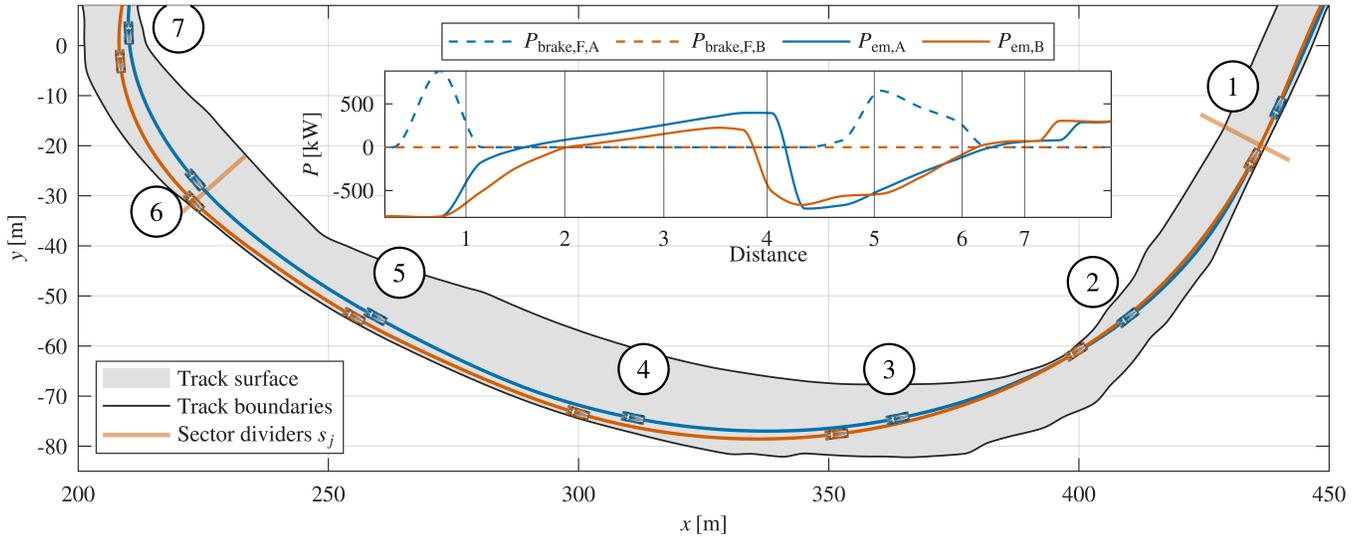}
	\caption{Driven racing lines, electric motor power, and front braking power during the overtaking maneuver through the combination of Turns~9 and~10. Agent~A brakes more aggressively into Turn~9~(2), resulting in a reduced entry velocity that enables earlier acceleration~(3) and a tighter exit line~(5). By spending additional energy, agent~A moves ahead of its opponent at the next sector divider~(7), thereby gaining the right to select the preferred racing line and complete the overtake~(8). The cars and track are shown at their true proportions.}
	\label{fig:overtake}
\end{figure*}

A detailed view of the overtaking maneuver is provided in Fig.~\ref{fig:overtake}, which depicts the trajectories of both agents. Agent~A deviates from the trajectory of agent~B under braking into Turn~9 in order to maintain aerodynamic downforce and therefore grip. At the expense of reduced regenerative braking, agent~A applies the friction brakes more aggressively and adopts a wider entry line, enabling earlier acceleration at corner exit and allowing it to draw alongside its opponent. Approaching Turn~10, agent~A again sacrifices regenerative energy by braking later and more forcefully, emerging ahead by the sector divider. This grants agent~A the right to select its preferred trajectory into the corner, forcing agent~B to yield and follow. The success of this maneuver is enabled by the combined battery energy advantage arising from the higher initial allocation and the energy savings obtained through slipstreaming earlier in the lap.

A noteworthy observation is that none of the solutions exhibit aggressive defensive maneuvers by the leading vehicle. This outcome is consistent with the deterministic setting and the chosen objective, which is formulated as a weighted sum of final finishing position and lap time. When an overtake is inevitable, blocking maneuvers provide no strategic benefit, as they incur a significant lap time penalty without altering the final outcome. In the deterministic setting of the proposed framework, the leading agent therefore optimally refrains from defensive actions and instead minimizes lap time, accepting the loss of position when it cannot be prevented.

\subsection{Multi-Lap Multi-Agent}
\begin{figure}[t]
	\centering
	\includegraphics[width=1\linewidth]{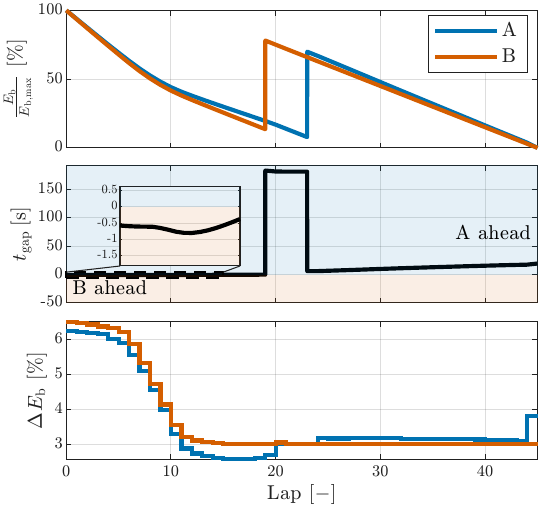}
	\caption{The evolution of the battery energy, time gap and the battery energy allocation per lap for a 45-lap race using the obtained policy. Both agents start with a full battery and an initial time gap of -0.5\unit[]{s} in favor of agent~B. agent A is able to save energy by staying in the slipstream of agent B. This results in a shorter charging time during the pit stop, resulting in agent A winning the race.}
	\label{fig:Strategy}
\end{figure}

This section presents the obtained policy for determining the battery energy allocation, battery energy pit threshold and the amount of energy to recharge the battery with.
We select a 45 lap race, the initial time gap is set to -0.5\unit{s}, which corresponds to 4 car lengths, and both agents start with a full battery. We assume a \emph{rolling start}, which is the standard in endurance racing. The minimum and maximum amount of battery energy allocation of the leader was set to 3\% and 9\%, respectively. Fig.~\ref{fig:Strategy} shows the evolution of the states and actions throughout the race.

Agent~A wins the race by conserving energy prior to the pit stop through sustained slipstreaming, while maintaining a time gap of approximately half a second. This energy saving reduces the required charging duration during the pit stop, allowing agent~A to exit the pit lane 6.0\unit{s} ahead of agent~B. After the pit stop, agent~A has 4.0\% more battery energy and the same number of laps to complete, allowing it to extend the gap to 19.4\unit{s} during the remaining laps of the race. The gap is sufficiently large to prevent agent~B from benefiting from slipstream-induced energy savings, thereby securing agent~A’s advantage to the end of the race.

Agent~B is unable to mount a successful defense against its opponent’s strategy, but nevertheless adopts several tactics to maximize its probability of winning the race. The first tactic is to push aggressively and expend a large amount of energy during the opening laps. This serves two purposes. First, higher driving speeds reduce the potential energy savings available to the following agent, thereby limiting the advantage agent~A can gain through slipstreaming. Second, the policy was trained against a pool of previous policy versions rather than a single fixed opponent. In this setting, aggressive early driving often induces the opponent to drop back in order to conserve energy, increasing the time gap and eliminating any benefit from slipstreaming.

The second tactic is to enter the pit lane as early as possible to break the slipstream provided to the opponent. Since charging power decreases significantly above a battery state of charge of 80\%, Agent~B delays its pit stop until the remaining race distance can be completed with a post-charge energy level of 78.1\%. This strategy minimizes both the charging duration and the time spent providing a slipstream to its opponent. Once agent~B has pitted, the exact timing of agent~A’s pit stop becomes largely irrelevant, as agent~A can simply pit a few laps later and charge sufficient energy to complete the remaining laps.

The final policy was evaluated for 1000 races and the agent that started the race in second place always won. This is in contrast to single-agent policies as their strategy would be identical for both agents, resulting in the finishing positions being identical to the starting positions. This result highlights the importance of competitor-aware race policies.

\section{Conclusion}\label{sec:Conclusion}
This paper presents a bi-level framework for competitor-aware race management in electric endurance racing. At the lower level, we use game-theoretic methods to determine the optimal driver inputs by solving a multi-agent optimal control problem in which each agent tries to finish ahead of their opponent, while respecting asymmetric collision-avoidance constraints inspired by real motorsport rules. At the upper level, we model the race across multiple laps using the solutions of the single lap problem, and train a reinforcement learning policy to determine per-lap energy allocation and charging decisions that maximize the probability of winning. At the single-lap level, we showed that our method produces realistic competitive trajectories in which the agents adhere to real-world racing principles. Our results show that aerodynamic slipstreaming is a powerful strategic tool. At the single-lap-level the following agent exploits reduced drag to recover energy and execute overtakes, and at the race-level the same mechanism lets an agent accumulate an energy buffer that shortens charging time and allows them to win the race. Defending agents therefore adopt countermeasures by driving more aggressively to deny the draft and pitting earlier to break sustained slipstreaming, trading battery energy and longer charge times for reduced advantage to the opponent. In future work we would like to train the agent for races which require multiple charging stops, and include more competitors.

%
%

\bibliographystyle{IEEEtran}
\bibliography{Bibliography}
\end{document}